\documentstyle[12pt]{article}
\begin{document}

\def\ds{\displaystyle}
\def\beq{\begin{equation}}
\def\eeq{\end{equation}}
\def\bea{\begin{eqnarray}}
\def\eea{\end{eqnarray}}
\def\beeq{\begin{eqnarray}}
\def\eeeq{\end{eqnarray}}
\def\ve{\vert}
\def\vel{\left|}
\def\bpll{B\rar\pi \ell^+ \ell^-}
\def\ver{\right|}
\def\nnb{\nonumber}
\def\ga{\left(}
\def\dr{\right)}
\def\aga{\left\{}
\def\adr{\right\}}
\def\lla{\left<}
\def\rra{\right>}
\def\rar{\rightarrow}
\def\nnb{\nonumber}
\def\la{\langle}
\def\ra{\rangle}
\def\ba{\begin{array}}
\def\ea{\end{array}}
\def\tr{\mbox{Tr}}
\def\ssp{{\Sigma^{*+}}}
\def\sso{{\Sigma^{*0}}}
\def\ssm{{\Sigma^{*-}}}
\def\xis0{{\Xi^{*0}}}
\def\xism{{\Xi^{*-}}}
\def\qs{\la \bar s s \ra}
\def\qu{\la \bar u u \ra}
\def\qd{\la \bar d d \ra}
\def\qq{\la \bar q q \ra}
\def\gGgG{\la g^2 G^2 \ra}
\def\q{\gamma_5 \not\!q}
\def\x{\gamma_5 \not\!x}
\def\g5{\gamma_5}
\def\sb{S_Q^{cf}}
\def\sd{S_d^{be}}
\def\su{S_u^{ad}}
\def\ss{S_s^{??}}
\def\ll{\Lambda}
\def\lb{\Lambda_b}
\def\sbp{{S}_Q^{'cf}}
\def\sdp{{S}_d^{'be}}
\def\sup{{S}_u^{'ad}}
\def\ssp{{S}_s^{'??}}
\def\sig{\sigma_{\mu \nu} \gamma_5 p^\mu q^\nu}
\def\fo{f_0(\frac{s_0}{M^2})}
\def\ffi{f_1(\frac{s_0}{M^2})}
\def\fii{f_2(\frac{s_0}{M^2})}
\def\O{{\cal O}}
\def\sl{{\Sigma^0 \Lambda}}
\def\es{\!\!\! &=& \!\!\!}
\def\ar{&+& \!\!\!}
\def\ek{&-& \!\!\!}
\def\cp{&\times& \!\!\!}
\def\se{\!\!\! &\simeq& \!\!\!}
\def\hml{\hat{m}_{\ell}}
\def\rr{\hat{r}_{\Lambda}}
\def\ss{\hat{s}}

% ...........................................................

\renewcommand{\textfraction}{0.2}    %float (figures) parameters
\renewcommand{\topfraction}{0.8}

\renewcommand{\bottomfraction}{0.4}
\renewcommand{\floatpagefraction}{0.8}
\newcommand\mysection{\setcounter{equation}{0}\section}

\def\baeq{\begin{appeq}}     \def\eaeq{\end{appeq}}
\def\baeeq{\begin{appeeq}}   \def\eaeeq{\end{appeeq}}
\newenvironment{appeq}{\beq}{\eeq}
\newenvironment{appeeq}{\beeq}{\eeeq}
\def\bAPP#1#2{
 \markright{APPENDIX #1}
 \addcontentsline{toc}{section}{Appendix #1: #2}
 \medskip
 \medskip
 \begin{center}      {\bf\LARGE Appendix #1 :}{\quad\Large\bf #2}
% \begin{center}      {\bf\LARGE Appendix  :}{\quad\Large\bf #2}
\end{center}
 \renewcommand{\thesection}{#1.\arabic{section}}
\setcounter{equation}{0}
        \renewcommand{\thehran}{#1.\arabic{hran}}
\renewenvironment{appeq}
  {  \renewcommand{\theequation}{#1.\arabic{equation}}
     \beq
  }{\eeq}
\renewenvironment{appeeq}
  {  \renewcommand{\theequation}{#1.\arabic{equation}}
     \beeq
  }{\eeeq}
\nopagebreak \noindent}

\def\eAPP{\renewcommand{\thehran}{\thesection.\arabic{hran}}}

\renewcommand{\theequation}{\arabic{equation}}
\newcounter{hran}
\renewcommand{\thehran}{\thesection.\arabic{hran}}

\def\bmini{\setcounter{hran}{\value{equation}}
\refstepcounter{hran}\setcounter{equation}{0}
\renewcommand{\theequation}{\thehran\alph{equation}}\begin{eqnarray}}
\def\bminiG#1{\setcounter{hran}{\value{equation}}
\refstepcounter{hran}\setcounter{equation}{-1}
\renewcommand{\theequation}{\thehran\alph{equation}}
\refstepcounter{equation}\label{#1}\begin{eqnarray}}

%       the stuff below defines \eqalign and \eqalignno in such a
%       way that they will run on Latex

\newskip\humongous \humongous=0pt plus 1000pt minus 1000pt
\def\caja{\mathsurround=0pt}
%\def\eqalign#1{\,\vcenter{\openup1\jot
%\caja   %\ialign{\strut \hfil$\displaystyle{##}$&$
%\displaystyle{{}##}$\hfil\crcr#1\crcr}
%}\,}

% ...........................................................

\title{
 {\small \begin{flushright}
IPM/P-2007/000\\
\today
\end{flushright}}
       {\Large
                 {\bf
 Analysis of the  $B \rar \pi \ell^+ \ell^-$ Decay in the Standard Model with Fourth Generation.
                 }
         }
      }

\author{\vspace{1cm}\\
{\small  V.Bashiry$^1$\thanks {e-mail: bashiry@ipm.ir}\,\,, K.
Zeynali$^2$\thanks {e-mail: k.zeinali@arums.ac.ir}\,\,,
} \\
{\small $^1$ Institute for Studies in Theoretical Physics and
Mathematics (IPM),}\\{ \small  P.O. Box 19395-5531, Tehran, Iran }\\
{\small $^2$ Department of Sciences, Faculty of Medicine, Ardabil
University of Medical Sciences,}\\ {\small Ardabil, Iran }\\}
\date{}
\begin{titlepage}
\maketitle
\thispagestyle{empty}

\begin{abstract}
We investigate the influence of the fourth generation quarks on
the branching ratio and the CP-asymmetry  in $B \rar \pi \ell^+
\ell^-$ decay. Taking the $|V_{t'd}V_{t'b}|\sim 0.001$ with phase
about $10^\circ$, which is consistent with the $sin2\phi_1$ of the
CKM and the $B_d$ mixing parameter $\Delta m_{B_d}$, we obtain
that for both ($\mu, \, \tau$) channels the branching ratio, the
magnitude of CP-asymmetry and lepton polarization depict strong
dependency on the 4th generation quarks mass and mixing
parameters. These results  can serve as a good tool to search for
new physics effects, precisely, to search for the fourth
generation quarks($t',\, b')$ via its indirect manifestations in
loop diagrams.
\end{abstract}

%\vspace{1cm}
~~~PACS numbers: 12.60.--i, 13.30.--a, 14.20.Mr
\end{titlepage}

\section{Introduction}
The first evidence of new strong interactions may be a
sufficiently massive fourth family observed at the LHC. The fourth
family masses, of the leptons in particular, are constrained by
the electroweak precision data, and this leads to signatures at
the LHC that may imply early discovery. The discovery of a fourth
family could potentially come quite early. The fourth family
quarks and leptons are free to have mass mixing (CKM mixing) with
the lighter fermions, and thus treelevel charged-current decays or
some loop induced(FCNC) decays may affected by the existence of
the 4th generation top like quark($t'$) (i.e, see
\cite{Bashiry2007}-\cite{Arhrib:2002md}). We will discuss
processes of this type that should be quite accessible at the LHC.
There are constraints on a fourth family\cite{Sultansoy:2000dm}.
From the strong constraint on the number of light neutrinos, we
know that the fourth family neutrino is heavy. The $S$ and $\rho$
parameters are sensitive to a fourth family, but the experimental
limits on these parameters have been evolving over the years in
such a way that the constraint on a fourth family has lowered. In
addition, the masses of the fourth family leptons may be such as
to produce negative S and T. As discussed in \cite{4thgood} and
the reference therein the constraints from S and T do not prohibit
the fourth family, but instead serve only to constrain the mass
spectrum of the fourth family quarks and leptons. The implied
masses for the fourth family leptons should make them particularly
accessible at the LHC, with neutrino pair production providing the
most interesting signatures. Any way, a sequential fourth family
is theoretically attractive because it makes it possible that a
theory of flavor is related to the breakdown of a simple family
gauge symmetry\cite{rev4th}. In contrast, new fermions not having
standard model quantum numbers would be more surprising and
difficult to understand.

 New Physics (NP) can be searched for in two
ways: either by raising the available energy at colliders to
produce new particles and reveal them directly, or by increasing
the experimental precision on certain processes involving Standard
Model (SM) particles as external states. The latter option,
indirect search for NP, should be pursued using processes that are
forbidden, very rare or precisely calculable in the SM. In this
respect, Flavor Changing Neutral Current (FCNC) and CP-violating
processes are among the most powerful probes of NP, since in the
SM they cannot arise at the tree-level and even at the loop level
they are strongly suppressed by the GIM mechanism. Furthermore, in
the quark sector they are all calculable in terms of the CKM
matrix, and in particular of the parameters $\bar \rho$ and $\bar
\eta$ in the generalized Wolfenstein
parametrization~\cite{Wolfenstein:1983yz}. Unfortunately, in many
cases a deep understanding of hadronic dynamics is required in
order to be able to extract the relevant short-distance
information from measured processes. Lattice QCD and QCD sum rules
allow us to compute the necessary hadronic parameters in many
processes.  Indeed, the Unitarity Triangle Analysis (UTA) with
Lattice QCD input is extremely successful in determining $\bar
\rho$ and $\bar \eta$ and in constraining NP
contributions~\cite{hep-ph/0501199,hep-ph/0606167,hep-ph/0509219,hep-ph/0605213,hep-ph/0406184}.

Once the CKM matrix is precisely determined by means of the UTA,
it is possible to search for NP contributions. FCNC and
CP-violating are indeed the most sensitive probes of NP
contributions to penguin operators. Rare decays, induced by flavor
changing neutral current (FCNC) $b \rar s(d)$ transitions is at
the forefront of our quest to understand flavor and the origins of
CPV, offering one of the best probes for New Physics (NP) beyond
the Standard Model (SM) \cite{Willey}--\cite{Buras1995}. In
addition, there are important QCD corrections, which have recently
been calculated in the NNLL\cite{NNLL}. Moreover, $b \rar s(d)
\ell^+ \ell^-$ decay is also very sensitive to the new physics
beyond SM. New physics effects manifest themselves in rare decays
in two different ways, either through new combinations to the
Wilson coefficients or through the new structure of the operator
in the effective Hamiltonian, which is absent in the SM. A crucial
problem in the new physics search within flavor physics is the
optimal separation of new physics effects from uncertainties. It
is well known that inclusive decay modes are dominated partonic
contributions; non--perturbative corrections are in general rather
small\cite{Hurth}. Also, ratios of exclusive decay modes such as
asymmetries for $B\rar K(~K^\ast,~\rho,~\gamma)~ \ell^+ \ell^-$
decay \cite{R4621}--\cite{bashirychin} are well studied for
new--physics search. Here large parts of the hadronic
uncertainties, partially, cancel out.

In this paper, we investigate the possibility of searching for new
physics in the  $\bpll$ decay using the SM with four generations
of quarks($b',\, t'$). The fourth quark ($t'$), like $u,c,t$
quarks, contributes in the $b \rar s(d) $ transition at loop
level. It would clearly change the branching ratio and
CP-asymmetry. Note that, fourth generation effects on the
branching ratio have been widely studied in baryonic and
semileptonic $b\rar s$ transition
\cite{Hou:2006jy}--\cite{London:1989vf}. But, there isn't any
study related to the $b\rar d $ transitions .

The sensitivity of the physical abservable to the existence of
fourth generation quarks in the $B \rar K^\ast \ell^+ \ell^-$
decay is investigated in \cite{Arhrib:2002md} and it is obtained
that the CP asymmetry is very sensitive to the fourth generation
parameters ($m_{t'}$, $V_{t'b}V^*_{t'd}$ ). In this connection, it
is natural to ask whether the branching ratio, CP-asymmetry and
lepton polarization in $\bpll$ are sensitive to the fourth
generation parameters in the same way. In the present work, we try
to answer to these questions.

The paper is organized as follows: In section 2, using the
effective hamiltonian, the general expressions for the matrix
element and CP asymmetry of $ B \rar \pi \ell^+ \ell^-$  decay is
derived. Section 3 is devoted to calculations of lepton
polarization. In section 4, we investigate the sensitivity of the
above mentioned physical observable to the fourth generation
parameters ($m_{t'}$, $V_{t'b}V^*_{t'd}$ ).

\section{Matrix Element, Differential Decay Rate and CP Asymmetry}
With a sequential fourth generation, the Wilson coefficients
$C_7,\, C_9$ and $C_{10}$ receive contributions from the $t'$
quark loop, which we will denote as $C^{new}_{7,9,10}$ . Because a
sequential fourth generation couples in a similar way to the
photon and W, the effective Hamiltonian relevant for $b \rar d
\ell^+ \ell^-$ decay has the following formula:
 \bea\label{Hgen} {\cal H}_{eff} &=& \frac{4 G_F}{\sqrt{2}}
V_{tb}V_{td}^\ast \sum_{i=1}^{10} {\cal C}_i(\mu) \, {\cal
O}_i(\mu)~, \eea where the full set of the operators ${\cal
O}_i(\mu)$ and the corresponding expressions for the Wilson
coefficients ${\cal C}_i(\mu)$ in the SM are given in
\cite{R23}--\cite{R24}. As it has already been noted , the fourth
generation up type quark $t'$ is introduced in the same way as
$u,\ c,\ t$ quarks introduce in the SM, so, new operators do not
appear and clearly the full operator set is exactly the same as in
SM. The fourth generation changes the values of the Wilson
coefficients $C_7(\mu),~C_9(\mu)$ and $C_{10}(\mu)$, via virtual
exchange of the fourth generation up type quark $t^\prime$. The
above mentioned Wilson coefficients will explicitly change
  as
\bea\lambda_t C_i \rightarrow \lambda_t C^{SM}_i+\lambda_{t'}
C^{new}_i~,\eea where $\lambda_f=V_{f b}^\ast V_{fd}$. The unitarity
of the $4\times4$ CKM matrix leads to
\bea\lambda_u+\lambda_c+\lambda_t+\lambda_{t'}=0.\eea\ It follows
that \bea \lambda_t C^{SM}_i+\lambda_{t'} C^{new}_i=\lambda_c
C^{SM}_i+\lambda_{t'} (C^{new}_i-C^{SM}_i )\eea It is clear that,
for the $m_{t'}\rar m_t$ or $\lambda_{t'}\rar 0$, $\lambda_{t'}
(C^{new}_i-C^{SM}_i )$ term vanishes, as required by the GIM
mechanism. One can also write $C_i$'s in the following form
\bea\label{c4} C_7^{tot}(\mu) &=& C_7^{SM}(\mu) +
\frac{\lambda_{t'}}
{\lambda_t} C_7^{new} (\mu)~, \nnb \\
C_9^{tot}(\mu) &=& C_9^{SM}(\mu) +  \frac{\lambda_{t'}}
{\lambda_t}C_9^{new} (\mu) ~, \nnb \\
C_{10}^{tot}(\mu) &=& C_{10}^{SM}(\mu) +  \frac{\lambda_{t'}}
{\lambda_t} C_{10}^{new} (\mu)~, \eea where the last terms in
these expressions describe the contributions of the $t^\prime$
quark to the Wilson coefficients. $\lambda_{t'}$  can be
parameterized as: \bea {\label{parameter}}
\lambda_{t'}=V_{t^\prime b}^\ast V_{t^\prime
d}=r_{db}e^{i\phi_{db}}\eea

 In deriving Eq. (\ref{c4}), we factored
out the term $V_{tb}^\ast V_{td}$ in the effective Hamiltonian
given in Eq. (\ref{Hgen}). The explicit forms of the $C_i^{new}$
can be easily obtained from the corresponding expression of the
Wilson coefficients in SM by substituting $m_t \rar m_{t^\prime}$
(see \cite{R23,R25}). If the $d$ quark mass is neglected, the
above effective Hamiltonian leads to following matrix element for
the $b \rar d \ell^+ \ell^-$ decay \bea\label{e1} {\cal H}_{eff}
&=& \frac{G_F\alpha}{2\sqrt{2} \pi}
 V_{tb}V_{td}^\ast
\Bigg[ C_9^{tot} \, \bar d \gamma_\mu (1-\gamma_5) b \, \bar \ell
\gamma_\mu \ell + C_{10}^{tot} \bar d \gamma_\mu (1-\gamma_5) b \,
\bar \ell \gamma_\mu \gamma_5 \ell \nnb \\
&-& 2  C_7^{tot}\frac{m_b}{q^2} \bar d \sigma_{\mu\nu} q^\nu
(1+\gamma_5) b \, \bar \ell \gamma_\mu \ell \Bigg]~, \eea where
$q^2=(p_1+p_2)^2$ and $p_1$ and $p_2$ are the final leptons
four--momenta. The effective coefficient $C_9^{eff}$ can be
written in the following form: \bea C_9^{eff} =
\xi_1+\frac{\lambda_u}{\lambda_t}\xi_2 + Y(s')~, \eea where $s' =
q^2 / m_b^2$ and the function $Y(s')$ denotes the perturbative
part coming from one loop matrix elements of four quark operators
\cite{R23,R24}.  The explicit expressions for $\xi_1$, $\xi_2$,
and the values of $C_i$ in the SM can be found in \cite{R23,R24}.
\begin{table}
\renewcommand{\arraystretch}{1.5}
\addtolength{\arraycolsep}{3pt}
$$
\begin{array}{|c|c|c|c|c|c|c|c|c|}
\hline C_{1} & C_{2} & C_{3} & C_{4} & C_{5} & C_{6} & C_{7}^{SM} &
C_{9}^{SM} & C_{10}^{SM}\\ \hline
-0.248 & 1.107& 0.011& -0.026& 0.007& -0.031& -0.313& 4.344& -4.669\\
\hline
\end{array}
$$
\caption{The numerical values of the Wilson coefficients at $\mu =
m_{b}$ scale within the SM. The corresponding numerical value of
$C^{0}$ is $0.362$.}
\renewcommand{\arraystretch}{1}
\addtolength{\arraycolsep}{-3pt}
\end{table}

In addition to the short distance contribution, $Y_{per}(s')$
receives also long distance contributions, which have their origin
in the real $c\bar c$ and $u\bar u$ intermediate states. The
resonances are introduced by the Breit--Wigner distribution
through the replacement \cite{R26}--\cite{R28} \bea Y(s') =
Y_{per}(s') + \frac{3\pi}{\alpha^2} \, C^{(0)} \sum_{V_i=\psi_i}
\kappa_i \, \frac{m_{V_i} \Gamma(V_i \rar \ell^+ \ell^-)}
{m_{V_i}^2 - s' m_b^2 - i m_{V_i} \Gamma_{V_i}}~, \eea where
$C^{(0)}= 3 C_1 + C_2 + 3 C_3 + C_4 + 3 C_5 + C_6$. The
phenomenological parameters $\kappa_i$ can be fixed from
experimental measurements of  semileptonic B decays (i.e, ${\cal
B} (B \rar K^\ast V_i \rar K^\ast \ell^+ \ell^-) = {\cal B} (B
\rar K^\ast V_i)\, {\cal B} ( V_i \rar \ell^+ \ell^-)$, where the
data for the right hand side is given in \cite{R29}. For the
lowest resonances $J/\psi$ and $\psi^\prime$ one can use $\kappa =
1.65$ and $\kappa = 2.36$, respectively (see \cite{R30})).

One has to sandwich the inclusive effective hamiltonian between
initial hadron state $B(p_B)$ and final hadron state $\pi(p_{\pi})$
to obtain the matrix element for the exclusive decay $B (p_{B}) \to
\pi(p_{\pi})~ \ell^+ (p_+)\ell^-(p_-)$. It follows from Eq.
(\ref{e1}) that in order to calculate the decay width and other
physical observable of the exclusive $B \rar \pi \ell^+ \ell^-$
decay, the following  matrix elements in terms of form factors
\bea\label{12}
\langle\pi(p_\pi)|\bar{d}\gamma_\mu(1-\gamma^5)b|B(p_B)\rangle&=&f^+(q^2)(p_\pi+p_B)_\mu+f^-(q^2)q_\mu,
 \eea
\bea\label{13}
\langle\pi(p_\pi)|\bar{d}i\sigma_{\mu\nu}q^\nu(1+\gamma^5)b|B(p_B)\rangle&=&[q^2(p_\pi+p_B)_\mu-q_\mu(m_B^2-m_\pi^2)]f_\nu(q^2),
\eea
 have to be
calculated. In other words, the exclusive $B \rar \pi \ell^+
\ell^-$ decay which is described in terms of the matrix elements
of the quark operators given in Eq. (\ref{e1}) over meson states,
can be parameterized in terms of form factors ($f^+\,f^-$and $f_v$
). We observe that in calculating the physical observable at
hadronic level, we face the problem of computing the form factors.
This problem is related to the nonperturbative sector of QCD and
it can be solved only in framework a nonperturbative approach. In
the present work, we will use of the results the constituent quark
model predictions for the form factors.

 Now, we can obtain the matrix element which is as follows: \bea\label{matrix}
  M^{B\rightarrow\pi}&=&\frac{G_F\alpha}{2\sqrt{2}\pi}V_{tb}V_{td}^*\Bigg
  \{(2Ap_\pi^\mu+Bq^\mu)\bar{\ell}\gamma_\mu\ell+(2Gp_\pi^\mu+Dq^\mu)
  \bar{\ell}\gamma_\mu\gamma^5\ell\Bigg\},
 \eea
 where
 \bea\label{15}
A&=&C_9^{new}f^+-2m_BC_7^{new}f_v,
\\ \nnb
B&=&C_9^{new}(f^++f^-)+2\frac{m_B}{q^2}C_7^{new}f_v(m_B^2-m_\pi^2-q^2),
\\ \nnb
G&=&C_{10}^{new}f^+,
\\ \nnb
D&=&C_{10}^{new}(f^++f^-),
 \eea

From this expression of the matrix element, for the unpolarized
differential decay width we get the following result: \bea\label{16}
\Bigg(\frac{d\Gamma^\pi}{ds}\Bigg)_0&=&\frac{G_F^2\alpha^2}{2^{10}\pi^5}|V_{tb}V_{td}^*|^2m_B^3v\sqrt{\lambda_\pi}\Delta_\pi,
\eea

\bea\label{17}
 \Delta_\pi&=&\frac{1}{3}m_B^2\lambda_\pi(3-v^2)(|A|^2+|G|^2)+16m_\ell^2r_\pi|G|^2+4m_\ell^2s|D|^2
\\ \nnb &+&
8m_\ell^2(1-r_\pi-s)Re[GD^*],
 \eea
 with $r_\pi=m_\pi^2/m_B^2, \lambda_\pi=r_\pi^2+(s-1)^2-2r_\pi(s+1),
 v=\sqrt{1-\frac{4t^2}{s}}$ and
$t=m_\ell/m_B.$

Another physical quantity is normalized CP violating  asymmetry
which can be defined for both polarized and unpolarized leptons.
We aim to obtain normalized CP violating asymmetry for the
unpolarized leptons. The standard definition are given as: \bea
\label{acp} A^\pi_{CP}(\hat{s}) = \frac{\ds{\ga
\frac{d\Gamma^\pi}{d\hat{s}}\dr_0}- \ds{\ga
\frac{d\bar{\Gamma}^\pi}{d\hat{s}}\dr_0} } {\ds{\ga
\frac{d\Gamma^\pi}{d\hat{s}}\dr_0}+ \ds{\ga
\frac{d\bar{\Gamma}^\pi}{d\hat{s}}\dr_0} } = \frac{\Delta_\pi
-\bar{\Delta}_\pi}{\Delta_\pi +\bar{\Delta}_\pi}~, \eea where \bea
\frac{d\Gamma^\pi}{d\hat{s}} = \frac{d\Gamma^\pi(b\rar
d\ell^+\ell^-)}{d\hat{s}},~\mbox{\rm and},~
\frac{d\bar{\Gamma}^\pi}{d\hat{s}} = \frac{d\bar{\Gamma}^\pi(b\rar
d\ell^+\ell^-)}{d\hat{s}}~,\nnb \eea and
$(d\bar{\Gamma}^\pi/d\hat{s})_0$ can be obtained from
$(d\Gamma^\pi/d\hat{s})_0$ by making the replacement \bea
\label{e5714} C_9^{eff} = \xi_1+\lambda_u\xi_2 \rar
\bar{C}_9^{eff}= \xi_1+\lambda_u^\ast\xi_2~. \eea Using this
definition and the expression for $\Delta_\pi(\hat{s})$ the CP
violating asymmetry contributed from SM3 and new contribution from
SM4 are: \bea A^\pi_{CP}(\hat{s})=\frac{-\Sigma^{SM}-\Sigma^{new}}
{\Delta^1_{\pi}+\Sigma^{SM}+\Sigma^{new}}\eea where
\bea\label{AcpSM-kazem}
\Sigma^{SM}(\hat{s})&=&4Im{(\lambda_u)}\Bigg\{f^{+^2}Im{(\xi_1^*\xi_2)}+2f^+f_vm_B
Im{(C_7\xi_2^*)}\Bigg\}, \eea \bea\label{Acpnew-kazem}
 \Sigma^{new}(\hat{s})&=&4Im(\frac{\lambda_{t^\prime}}{\lambda_t})\Bigg\{2f^+f_vm_B[Im(c_7c_9^{new*})-
Im(c_7^{new}\xi_1^*) ]
 \\ \nnb &+&
 f^{+^2}Im(c_9^{new}\xi_1^*)\Bigg\}
\\ \nnb &+&
4Im(\frac{\lambda_{t^\prime}}{\lambda_t}\lambda_u)\Bigg\{2f^+f_vm_BIm(c_7\xi_2^*)+f^{+^2}Im(\xi_1^*\xi_2)\Bigg\}
\\ \nnb &+&
4Im(\frac{\lambda_{t^\prime}^*}{\lambda_t^*}\lambda_u)\Bigg\{-f^{+^2}Im(c_9^{new}\xi_2^*)
+2f^+f_vm_BIm(c_7^{new}\xi_2^*)\Bigg\}
\\ \nnb &+&
4Im(\lambda_u)\Bigg|\frac{\lambda_{t^\prime}}{\lambda_t}\Bigg|^2\Bigg\{f^{+^2}Im(c_9^{new}\xi_2)
-2f^+f_vm_BIm(c_7^{new}\xi_2^*)\Bigg\},
 \eea
 and
\bea\Delta_{\pi}^1=\frac{3\Delta_{\pi}}{m_B^2\lambda_\pi(3-v^2)}.
\eea From this expression, it is firstly easy to see that in the
$\lambda_{t'}\rightarrow 0$ the SM3 result can be obtained.
Secondly, when $m_{t'}\rightarrow m_t$ the result of the SM4
coincide withe the SM3 as it has to be seen(See figures), even if
it is not obvious from the expressions.
\section{Lepton polarization}
 In order to now calculate the
polarization asymmetries of the lepton defined in the effective
four fermion interaction of Eq. (\ref{matrix}), we must first
define the orthogonal vectors $S$ in the rest frame of $\ell^-$
(where its vector is the polarization vector of the lepton). Note
that, we use the subscripts $L$, $N$ and $T$ to correspond to the
leptons being polarized along the longitudinal, normal and
transverse directions, respectively.
\begin{eqnarray}
S^\mu_L & \equiv & (0, \mathbf{e}_{L}) ~=~ \left(0,
\frac{\mathbf{p}_-}{|\mathbf{p}_-|}
\right) , \nonumber \\
S^\mu_N & \equiv & (0, \mathbf{e}_{N}) ~=~ \left(0,
\frac{\mathbf{p_{\pi}} \times \mathbf{p}_-}{|\mathbf{p_{\pi}} \times
\mathbf{p}_- |}\right) , \nonumber \\
S^\mu_T & \equiv & (0, \mathbf{e}_{T}) ~=~ \left(0, \mathbf{e}_{N}
\times \mathbf{e}_{L}\right) , \label{sec3:eq:1} \eea

where $\mathbf{p}_-$ and $\mathbf{p_{\pi}}$ are the three momenta
of the $\ell^-$ and $\pi$ particles, respectively. The
longitudinal unit vectors is boosted to the CM frame of $\ell^-
\ell^+$ by Lorenz transformation:
\begin{eqnarray}
S^\mu_L & = & \left( \frac{|\mathbf{p}_-|}{m_\ell}, \frac{E_{\ell}
\mathbf{p}_-}{m_\ell |\mathbf{p}_-|} \right) , \nonumber
\end{eqnarray}
while the other two vectors remain unchanged. The polarization
asymmetries can now be calculated using the spin projector ${1 \over
2}(1 + \gamma_5 \!\!\not\!\! S)$ for $\ell^-$.

Provided the above expressions, we now define the single lepton
polarization. The definition of the polarized and normalized
differential decay rate is:
 \bea\label{diff}
\frac{d\Gamma^{\pi}(s,\vec{n})}{ds}=\frac{1}{2}\Bigg(\frac{d\Gamma^{\pi}}{ds}\Bigg)_0
[1+P^{\pi}_i \vec{e}.\vec{n}], \eea where a sume over $i=L,\,T,\,N$
is implied. Polarized components $P^{\pi}_i$ in Eq. (\ref{diff}) are
as follows:
  \bea \label{e6312} P_{i}^{\pi} =
\frac{ {d\Gamma^{\pi}(\vec{n}=\vec{e}_i)}{d\hat{s} } -
 {d\Gamma^{\pi}(\vec{n}=-\vec{e}_i)}/{d\hat{s} }} {{d\Gamma^{\pi}(\vec{n}=\vec{e}_i)}/{d\hat{s} } +
 {d\Gamma^{\pi}(\vec{n}=-\vec{e}_i)}/{{d\hat{s}}}}~, \eea
As a result, the different components of the $P^{\pi}_i$ are
given: \bea\label{pol} P^{\pi}_L&=&\frac{4m^2_B}{3\Delta_{\pi}}v
\lambda_{\pi}Re[A G^\ast],\\
P^{\pi}_T&=&\frac{m^2_B}{\sqrt{\hat{s}}\Delta_{\pi}}\pi
\sqrt{\lambda_{\pi}}t \Bigg(Re[A D^\ast]\hat{s}+Re[A G^\ast](1-r_\pi-\hat{s})\Bigg),\\
 P^{\pi}_N&=&0.\nnb\eea
A few words here are in order. Firstly, $P^{\pi}_N$ is zero in SM3
and SM4. It might be gained non--zero value in the case that the
type of the interaction change(i.e, scalar or tensor type
interactions may contribute). Secondly, $P^{\pi}_T$ is
proportional to the lepton mass and it is negligible for electron
case in SM3, considering SM4, it will be measurable in the case
that Wilson coefficients enhanced significantly by $m_{t'}$.

\section{Numerical analysis}
In this section, we will study the dependence of the total
branching ratio, CP asymmetry and lepton polarizations as well as
combined lepton polarization to the fourth quark mass($m_{t'}$)
and the product of quark mixing matrix elements ($V_{t^\prime
b}^\ast V_{t^\prime d}=r_{db}e^{i\phi_{db}}$). The main input
parameters in the calculations are the form factors. we have used
the results of the constituent quark model \cite{Melikov}, where
the form factors $f_T$ and $f_+$ can be parameterized as: \bea
f(q^2)=\frac{f(0)}{(1-q^2/T_f^2)[1-\sigma_1 q^2/M^2+\sigma_2
q^4/M^4]}\, . \eea In this model, $f_-$ is defined slightly
different and it is as: \bea f(q^2)=\frac{f(0)}{[1-\sigma_1
q^2/M^2+\sigma_2 q^4/M^4]}\, . \eea
\begin{table}[h]
\center
\begin{tabular}{|c c c c|}
\hline\hline
& $f(0)$ & $\sigma_1$ & $\sigma_2$ \\
\hline
$f_+$ & 0.29   & 0.48  &   \\
$F_0$ & 0.29  & 0.76   & 0.28  \\
$f_v$ & 0.28   & 0.48  &        \\
\hline\hline
\end{tabular}
\caption{$B\rar\pi$ transition form factors in the constituent quark
model.}\label{tabpi}
\end{table}
The parameters $f(0)$, $\sigma_i$'s can be found in Table
\ref{tabpi}.

The other input parameters used in our numerical analysis are as
follows:  \bea && m_B =5.28 \, \mbox{GeV} \, , \, m_b =4.8 \,
\mbox{GeV} \, , \,m_c=1.5 \, \mbox{GeV} \, , \, m_{\tau} =1.77 \,
\mbox{GeV} \, ,\,
 m_{e} =0.511 \, \mbox{MeV}, \nnb \\ &&m_{\mu} =0.105 \, \mbox{GeV},\, m_{\rho}=0.77 \,
  \mbox{GeV} \, , \, m_{d}=m_{u}= m_{\pi}=0.14 \, \mbox{GeV}  \, ,\,\nnb\\
&&|V_{cb}|=0.044\,,\,\alpha^{-1}=129\,,\,G_f=1.166\times10^{-5}\,
{\mbox{GeV}}^{-2}\,,\,\tau_B=1.56 \times 10^{-12}\,s\,. \eea
 In the
Wolfenstein parametrization of the CKM matrix
\cite{Wolfenstein:1983yz}, $\lambda_u$ is written as:
\bea\label{lamu}
\lambda_u=\frac{\rho(1-\rho)-\eta^2-i\eta}{(1-\rho)^2+\eta^2}+O(\lambda^2).
\eea Furthermore, we use the relation \bea\label{VtbVtd}
\frac{|V_{tb} V_{td}^\ast|^2}{|V_{cb}|^2} & = &
\lambda^2[(1-\rho)^2+\eta^2]+{\cal O}(\lambda^4) \eea where
$\lambda=\sin \theta_C\simeq 0.221$ and adopt the values of the
Wolfenstein parameters as $\rho=0.25$ and $\eta=0.34$.

 In order to perform quantitative analysis of the
total branching ratio, CP asymmetry and the lepton polarizations,
the values of the new parameters($m_{t'},\,r_{db},\,\phi_{db}$)
are needed. In the foregoing numerical analysis, we alter $m_{t'}$
in the range $175\le m_{t'} \le 600$GeV. The lower range is
because of the fact that the fourth generation up quark should be
heavier than the third ones($m_t \leq
m_{t'}$)\cite{Sultansoy:2000dm}. The upper range comes from the
experimental bounds on the $\rho$ and $S$ parameters of SM,
furthermore, a mass greater than the 600GeV will also contradict
with partial wave unitarity\cite{Sultansoy:2000dm}. As for mixing,
we use the result of Ref\cite{CKM4}where it is obtained that
$|V_{t'd}V_{t'b}|\sim 0.001$ with phase about $10^\circ$ is
consistent with the $sin2\phi_1$ of the CKM and the $B_d$ mixing
parameter $\Delta m_{B_d}$\cite{CKM4}.

Still, one more step can be proceeded. From explicit expressions of
the physical observable one can easily see that they depend on both
$\ss$ and the new parameters($m_{t'},\,r_{db}$). One may eliminate
the dependence of these quantities on one of the variables. We
eliminate the variable $\hat{s}$ by performing integration over
$\ss$ in the allowed kinematical region. The total branching ratio
and the averaged lepton polarizations  are defined as \bea
\label{ave}{\cal B}_r&=&\ds \int_{4
m_\ell^2/m_{B}^2}^{(1-\sqrt{\hat{r}_{\pi}})^2}
 \frac{d{\cal B}}{d\hat{s}} d\hat{s},
\nnb\\\lla P^{\pi}_{i}(A^{\pi}_{CP}) \rra &=& \frac{\ds \int_{4
m_\ell^2/m_{B}^2}^{(1-\sqrt{\hat{r}_{\pi}})^2}
P^{\pi}_i(A^{\pi}_{CP}) \frac{d{\cal B}}{d\hat{s}} d\hat{s}}
{{\cal{B}}_r}~. \eea

Figs. (1)--(8) depict the dependence of the total branching ratio,
unpolarized averaged CP asymmetry and averaged lepton polarization
for various  $r_{db}$ in terms of  $m_{t'}$. We should note, here,
that the dependency for various
$\phi_{db}\sim\{0^\circ-30^\circ\}$ is too weak, then we show the
results just for $\phi_{db}=15^\circ$. Looking at these figures,
the following outcomes are in order.

\begin{itemize}

\item ${\cal B}_r$  strongly depends on the fourth quark mass($m_{t'}$)
 and the product of quark mixing matrix elements($r_{db}$) for both $\mu$
 and $\tau$ channels. Furthermore, for both channels, ${\cal B}_r$ is an
 increasing function of both $m_{t'}$ and $r_{db}$.

\item $ P_L^\pi$ and $ A_{CP}^\pi$ are independent of the lepton
mass(See Eq. (\ref{pol}) and (\ref{acp})) as a result, for given
values of $\ss$ they are the same for $e,\,\mu,\,\mbox{and }\tau$
channels. The situation is different for the $\lla P_L^\pi\rra$
and $\lla A_{CP}^\pi\rra$, those values for $\tau$ channel are
less than as for $\mu$ and $e$ channel, because the phase integral
depends on the lepton mass($m_\ell$)(see Eq. (\ref{ave})). The SM3
value of $\lla P_L^\pi\rra$ and $\lla A_{CP}^\pi\rra$ are
negligible for the $\tau$ channel($\sim2\%$ and $\sim0.1\%$,
respectively). The SM4 suppress those approximately to zero. On
the other hand, $\lla P_L^\pi\rra$ and $\lla A_{CP}^\pi\rra$ for
$e,\,\mu$ channels are strongly depends to the SM4 parameters.
Moreover, their magnitudes are a decreasing function of the
$r_{db}$ and $m_{t'}$.

\item Although,  $\lla
P_T^\pi\rra$ strongly  depends on the fourth quark mass($m_{t'}$)
and the product of quark mixing matrix elements($r_{sb}$) for both
$\mu$ and $\tau$ channels. But, its magnitude is a decreasing
function of both $m_{t'}$ and $r_{sb}$. So, the existence of fourth
generation of quarks will suppress the magnitude of  $\lla
P_T^\pi\rra$.

\end{itemize}

In conclusion, we presented the systematic analysis of the $B \rar
\pi \ell^- \ell^+$ decay, by using the SM with fourth generation
of quarks. The sensitivity of the total branching ratio, CP
asymmetry and lepton polarization on the new parameters, coming
out of fourth generations, was studied. We found out that above
mentioned physical observable  depicted a strong dependence on the
fourth quark ($m_{t'}$) and the product of quark mixing matrix
elements ($V_{t^\prime b}^\ast V_{t^\prime
d}=r_{db}e^{i\phi_{db}}$). We obtained that the study of these
readily measurable quantities ,specially, for both $\mu$  case
could serve as a good tool to look for physics beyond the SM. More
precisely, the results could be used to indirect search to look
for fourth generation of quarks.

\section{Acknowledgment}
The authors would like to thank T. M. Aliev for his useful
discussions.

\newpage

\section*{Figure captions}

{\bf Fig. (1)} The dependence of the branching ratio of $B\rar \pi
\ell^-\ell^+$ where, $\ell=e,\,\mu$, on $m_{t'}$ for
 $r_{db}=0.001,\, 0.002,\,0.003$.\\ \\
{\bf Fig. (2)} The same as in Fig. (1), but for the $\tau$ lepton.\\ \\
{\bf Fig. (3)}  The dependence of the $\lla A_{CP}\rra$ on $m_{t'}$
for
 $r_{db}=0.001,\, 0.002,\,0.003$, where $\ell=e,\,\mu$.\\ \\
{\bf Fig. (4)} The dependence of the $\lla P_L\rra$ for $e$ lepton,
on $m_{t'}$ for
 $r_{db}=0.001,\, 0.002,\,0.003$.\\ \\
 {\bf Fig. (5)} The same as in Fig. (4), but for the $\mu$ lepton.\\
 \\
{\bf Fig. (6)}The same as in Fig. (4), but for the $\tau$
lepton.\\\\
 {\bf Fig. (7)}The dependence of the $\lla P_T\rra$ for $ \mu $ lepton,
on $m_{t'}$ for
 $r_{db}=0.001,\, 0.002,\,0.003$.\\ \\
{\bf Fig. (8)} The same as in Fig. (7), but for the $\tau$ lepton.\\
\\

\newpage
\begin{figure}
\vskip 1.5 cm
    \includegraphics{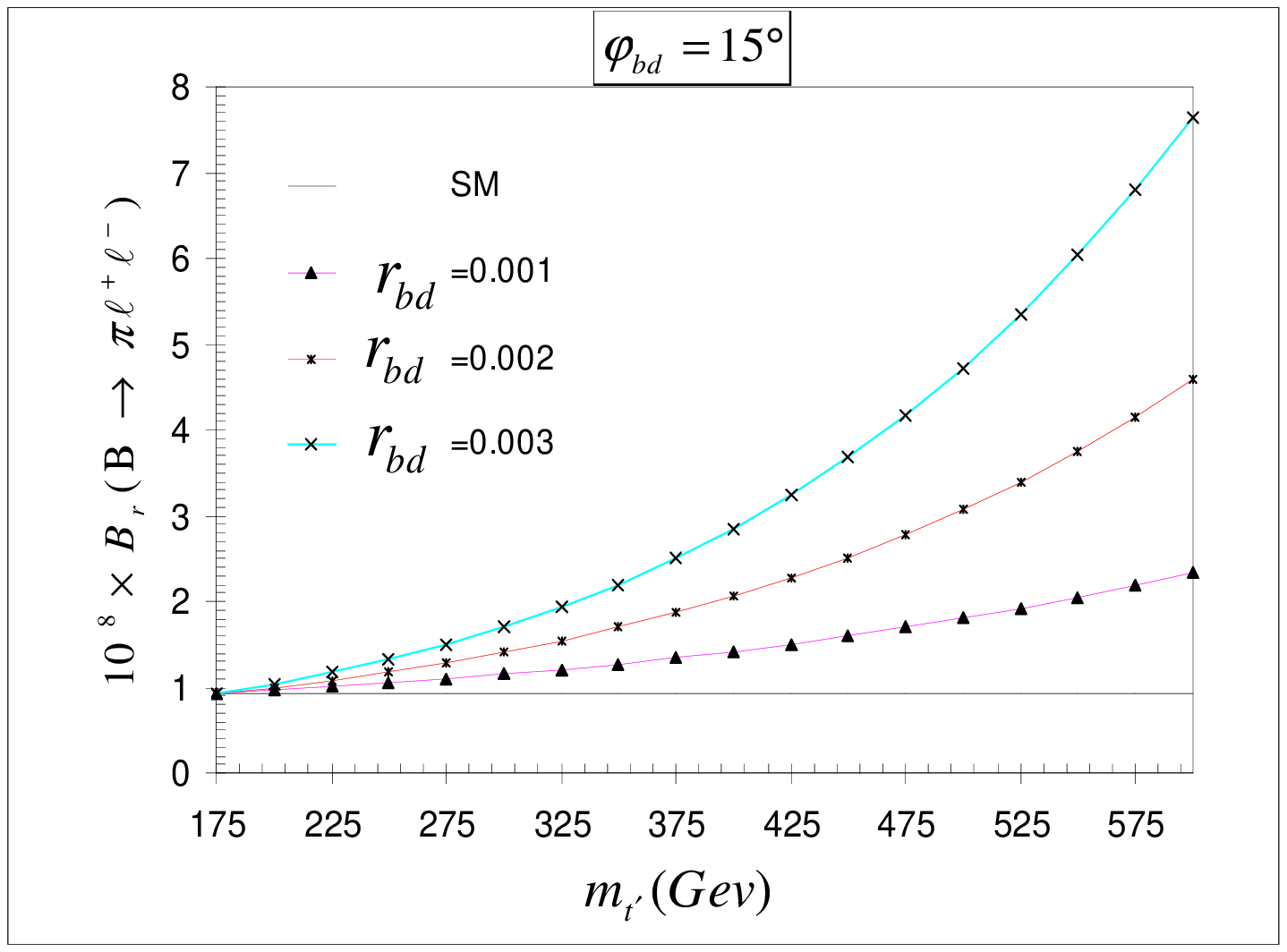}
\vskip 6.5cm \caption{}
\end{figure}
\begin{figure}
\vskip 1.5cm
    \includegraphics{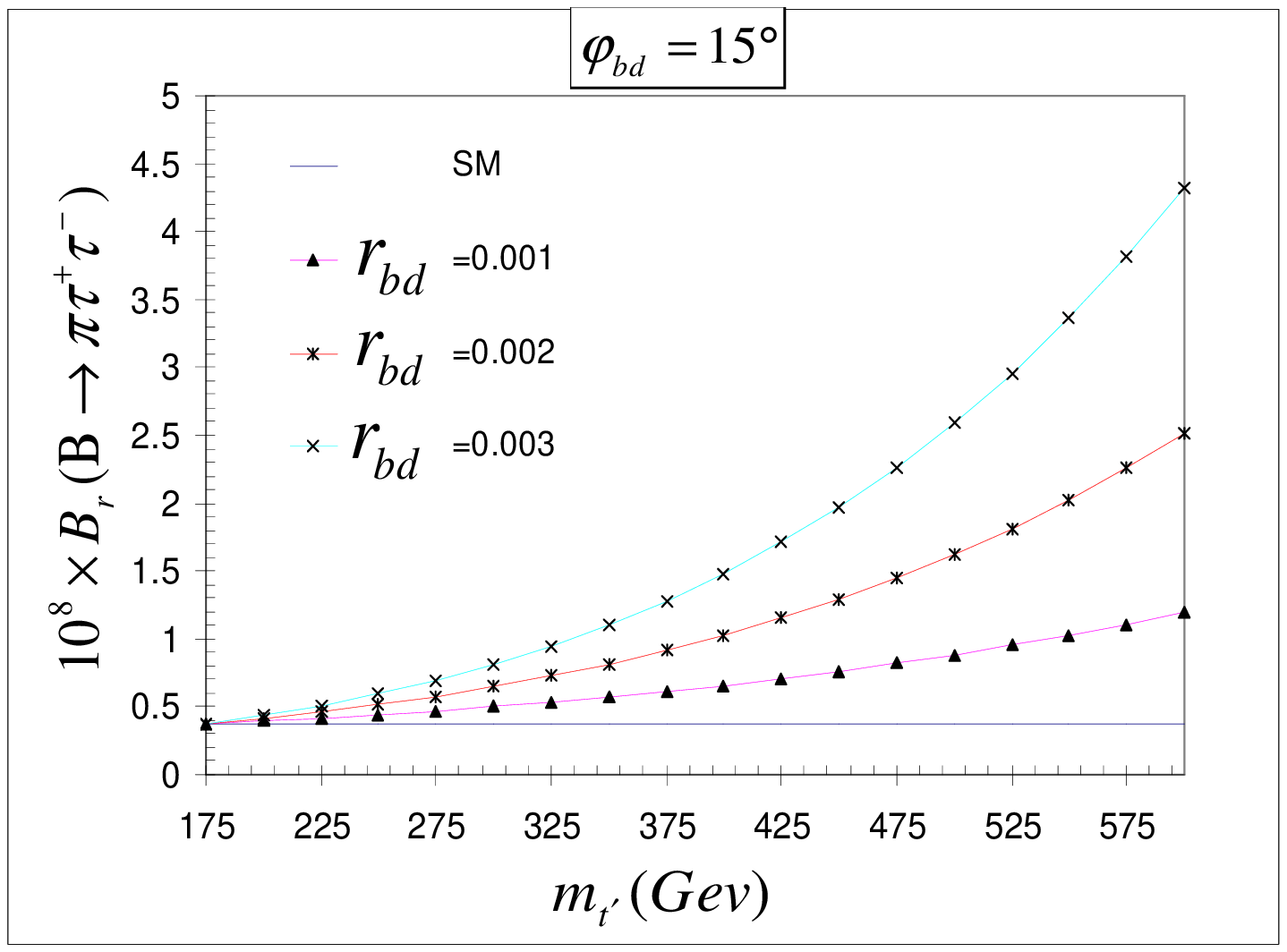}
\vskip 6.5cm \caption{}
\end{figure}
\begin{figure}
\vskip 1.5 cm
    \includegraphics{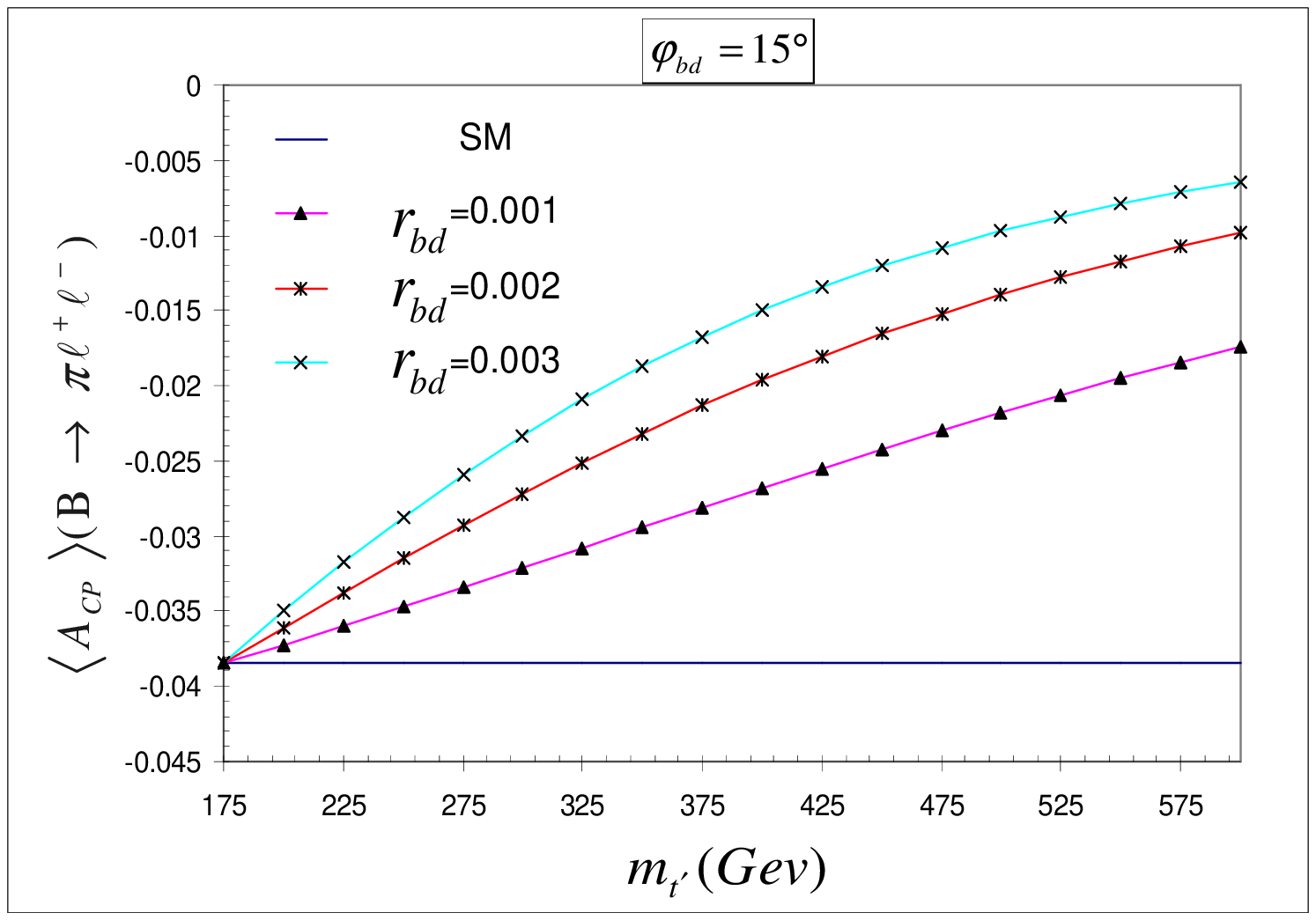}
\vskip 7.8cm \caption{}
\end{figure}

\begin{figure}
\vskip 1.5 cm
    \includegraphics{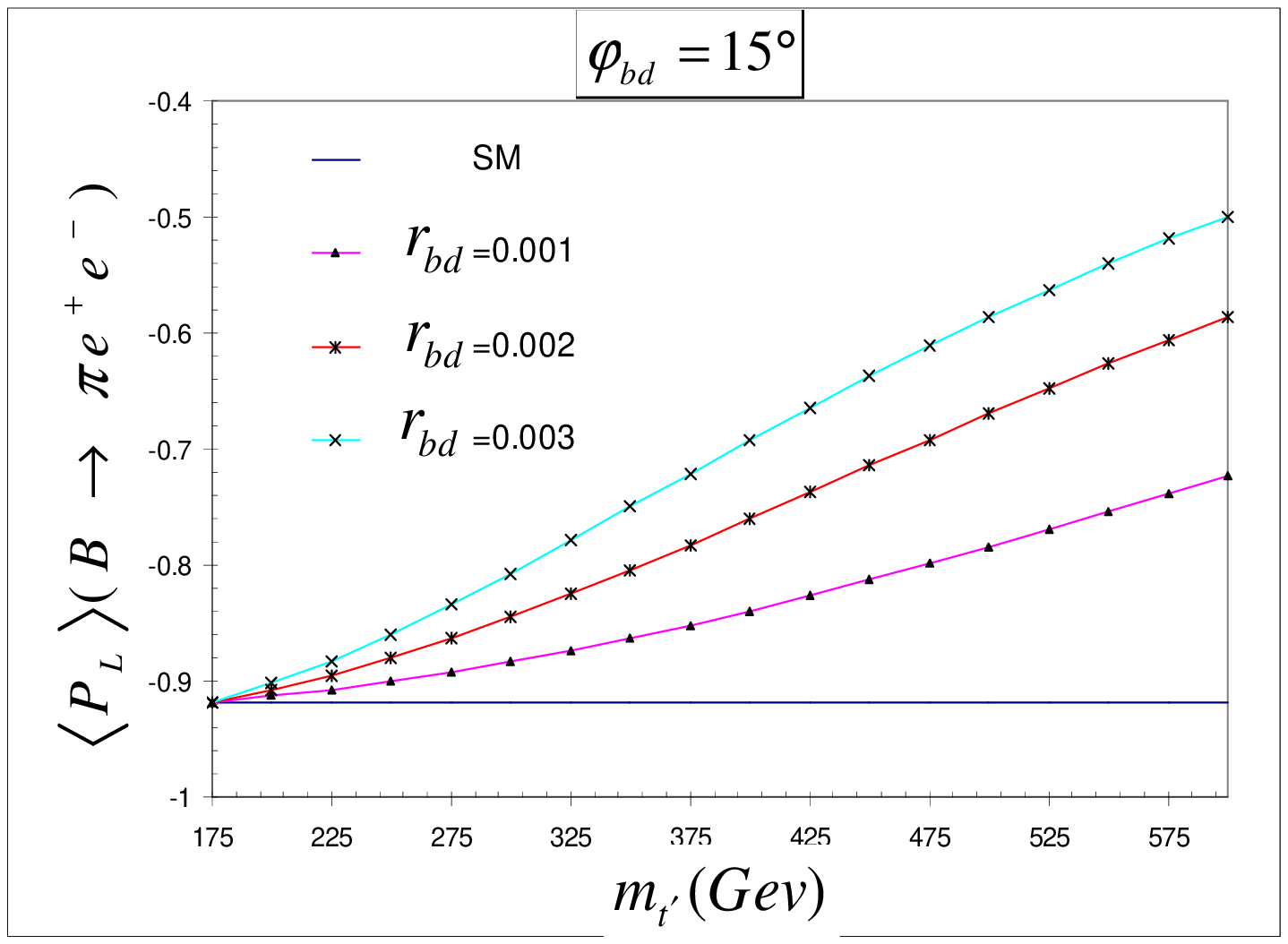}
\vskip 7cm \caption{}
\end{figure}
\begin{figure}
\vskip 1.5cm
    \includegraphics{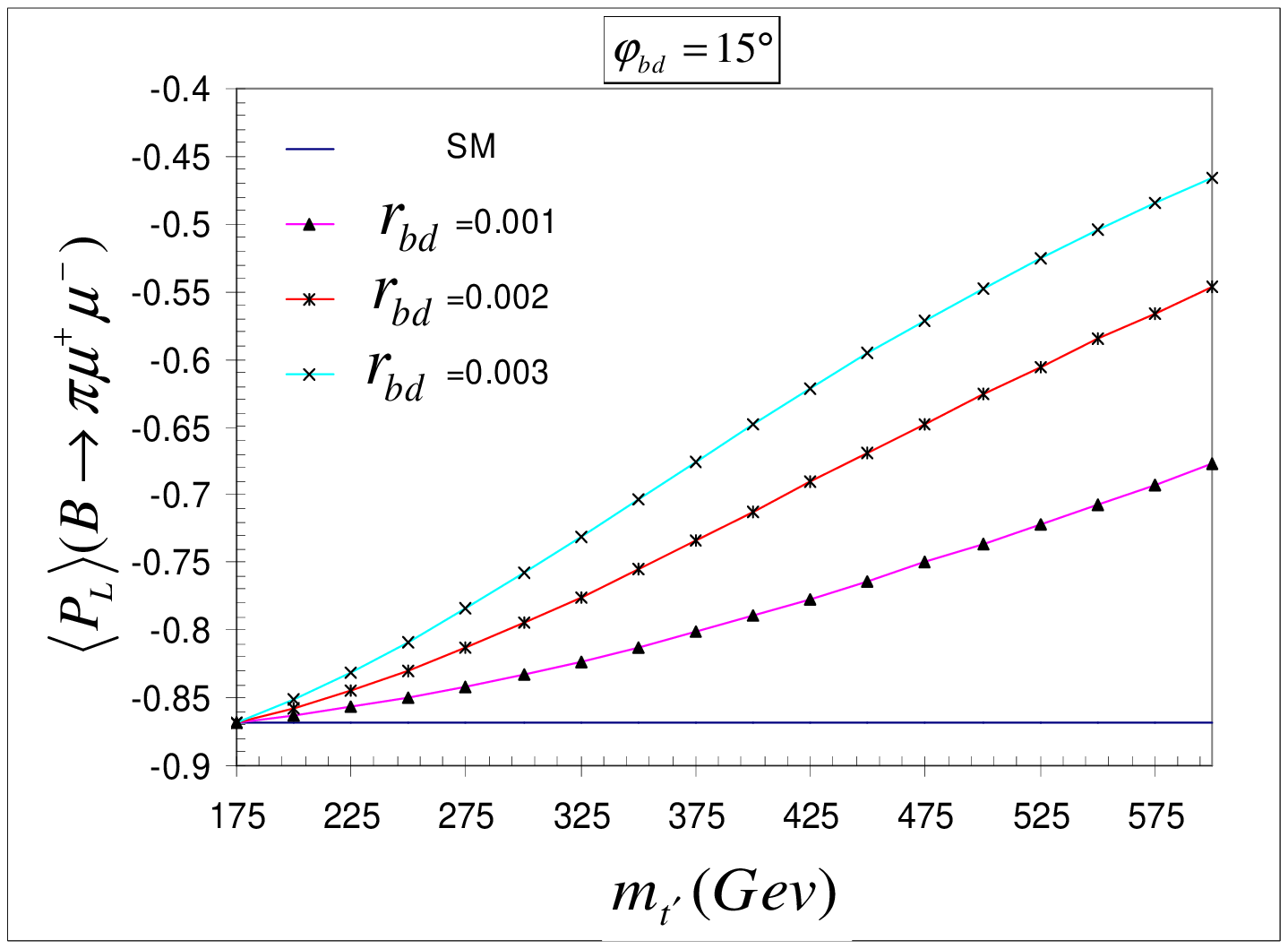}
\vskip 6.8cm \caption{}
\end{figure}
\begin{figure}
\vskip 1.5 cm
    \includegraphics{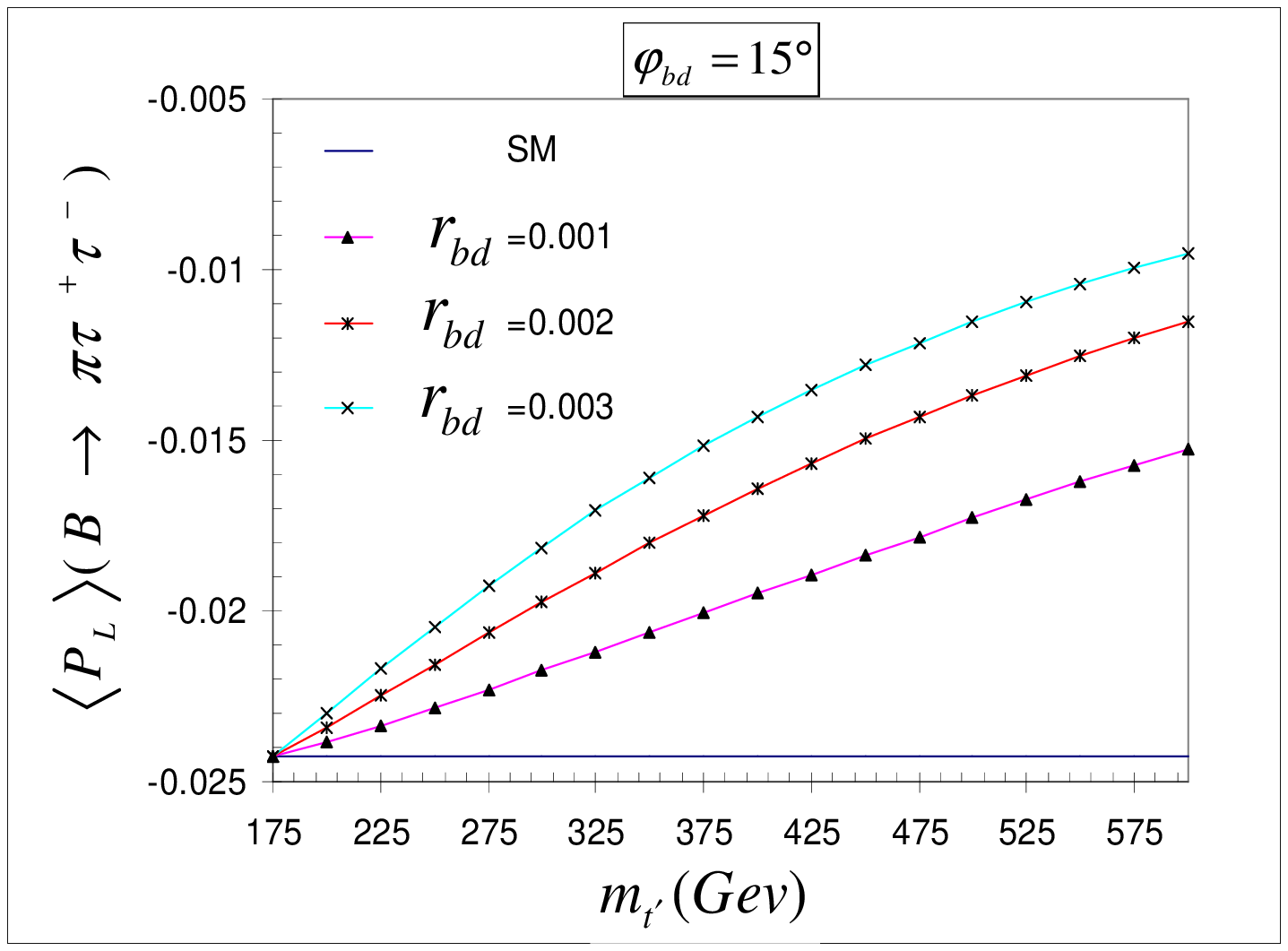}
\vskip 6.5cm \caption{}
\end{figure}
\begin{figure}
\vskip 1.5cm
    \includegraphics{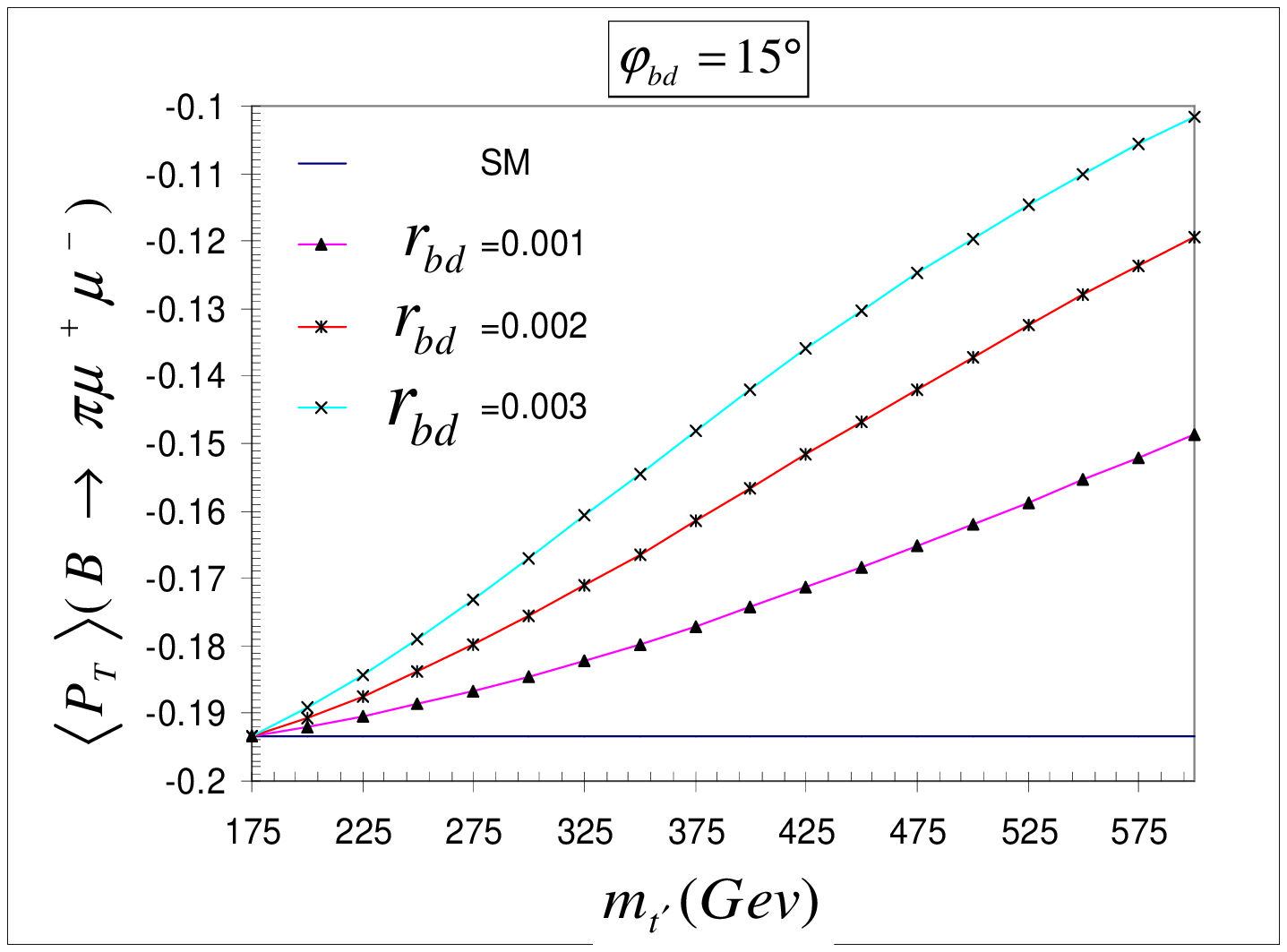}
\vskip 6.5cm \caption{}
\end{figure}
\begin{figure}
\vskip 1.5cm
    \includegraphics{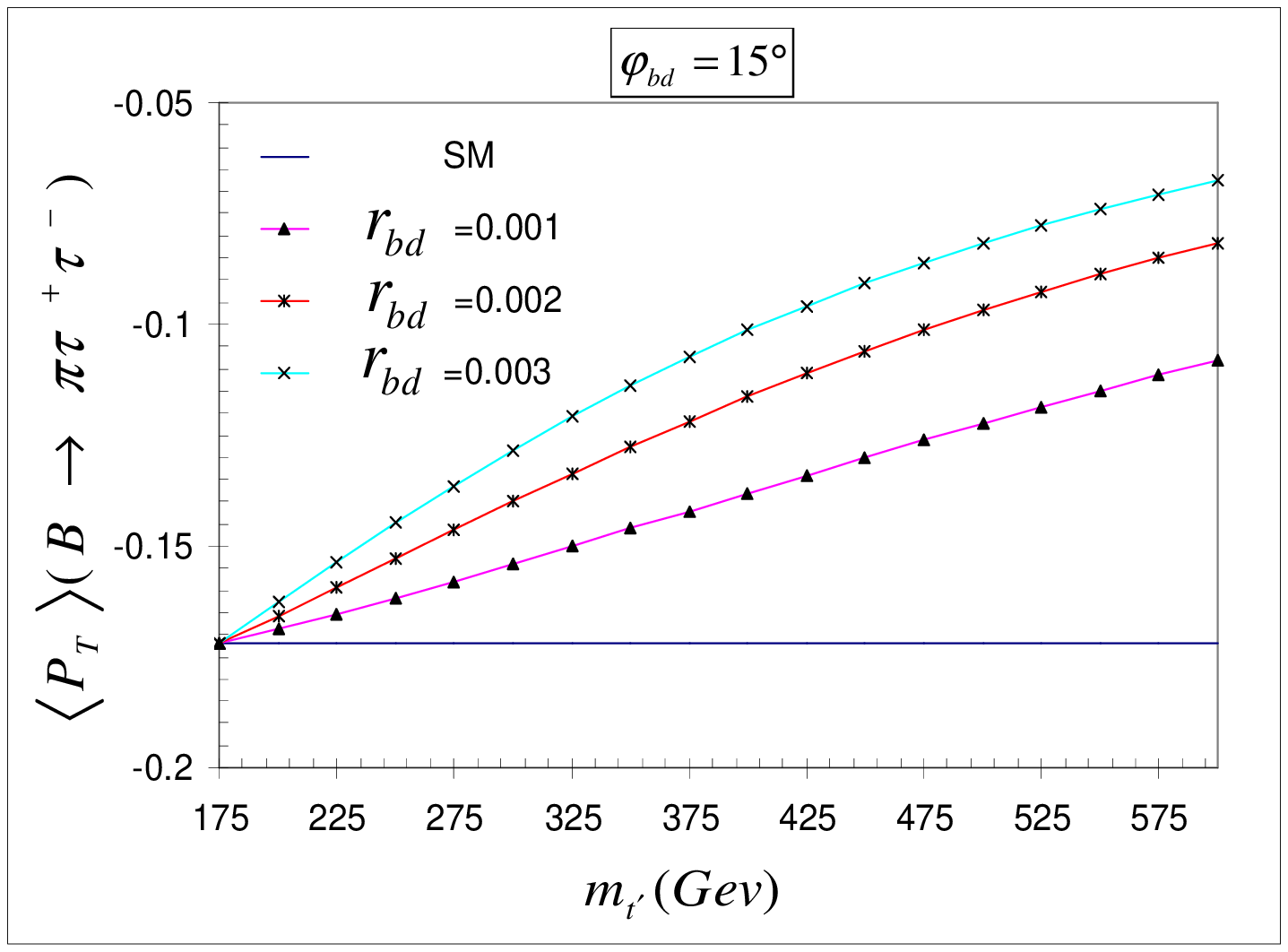}
\vskip 6.5cm \caption{}
\end{figure}

%\begin{figure}
%  % Requires \usepackage{graphicx}
%  \includegraphics[width=]{}\\
%  \caption{}\label{}
%\end{figure}

\end{document}